\documentclass[twocolumn]{aastex701}
\usepackage[utf8]{inputenc}
\usepackage{amsmath}

\definecolor{newred}{rgb}{0.1,0.6,0.75}
\definecolor{grey}{rgb}{0.2,0.27,0.57}
\definecolor{freeblue}{rgb}{0.2,0.25,0.45} 
\definecolor{freeblue2}{rgb}{0.2,0.25,0.4}
\definecolor{blue}{rgb}{0.45,0.41,0.98}
\definecolor{myblue}{rgb}{0.1,0.5,0.5}
\definecolor{guillaume}{rgb}{0.,0.5,0.65}
\definecolor{louloublue}{rgb}{0.2,0.2,0.65}
\definecolor{loulougreen}{rgb}{0.1,0.35,0.75}
\definecolor{violet}{rgb}{0.5,0.,0.5}

\newcolumntype{i}{>{\scriptsize}r}

\shortauthors{Readhead et al.}

\graphicspath{{./}{figs/}}

\begin{document}

\shorttitle{2:1 Harmonic Blazar Clocks}

\title{2:1 Harmonics in the Radio Light Curves of Supermassive Black Hole Binary Candidates}
\correspondingauthor{Anthony Readhead}
\email{acr@caltech.edu}

\author[0000-0001-9152-961X]{A.~C.~S.~Readhead}
\email{acr@caltech.edu}
\affiliation{Owens Valley Radio Observatory, California Institute of Technology,  Pasadena, CA 91125, USA}
\affiliation{Institute of Astrophysics, Foundation for Research and Technology-Hellas, GR-70013 Heraklion, Greece}

\author[0000-0003-2483-2103]{M. F. Aller}
\email{mfa@umich.edu}   
\affiliation{Department of Astronomy, University of Michigan, 323 West Hall, 1085 S. University Avenue, Ann Arbor, MI 48109, USA}

\author[0000-0002-1854-5506]{R. D. Blandford}
\email{rdb3@stanford.edu}   
\affiliation{Kavli Institute for Particle Astrophysics and Cosmology, Department of Physics,
Stanford University, Stanford, CA 94305, USA}

\author[0000-0002-6369-6266]{T. Surti}
\email{tsurti@caltech.edu}  
\affiliation{Owens Valley Radio Observatory, California Institute of Technology,  Pasadena, CA 91125, USA}

\author[0000-0001-5957-1412]{P. V.~De la Parra}
\email{phvergara@udec.cl}   
\affiliation{CePIA, Astronomy Department, Universidad de Concepci\'on,  Casilla 160-C, Concepci\'on, Chile}

\author[0000-0002-9545-7286]{A. G. Sullivan}
\email{andrew.sullivan@stanford.edu}   
\affiliation{Kavli Institute for Particle Astrophysics and Cosmology, Department of Physics,
Stanford University, Stanford, CA 94305, USA}
\affiliation{Columbia Astrophysics Laboratory, Columbia University in the City of New York, New York, NY 10027, USA}
\affiliation{Canadian Institute for Theoretical Astrophysics, University of Toronto, 60 St. George Street, Toronto, ON M5S 3H8, Canada}

\author[0000-0001-7016-1692]{P. Mr{\'o}z}
\email{Pmroz@astrouw.edu.pl}   
\affiliation{Astronomical Observatory, University of Warsaw, Al. Ujazdowskie 4, 00-478 Warszawa, Poland}

\author[0000-0002-8659-3729]{N. Murray}
\email{murray@cita.utoronto.ca} 
\affiliation{Canadian Institute for Theoretical Astrophysics, University of Toronto, 60 St. George Street, Toronto, ON M5S 3H8, Canada}

\author[0000-0003-0936-8488]{M. C.  Begelman}
\email{mitch@jila.colorado.edu}
\affiliation{JILA, University of Colorado and National Institute of Standards and Technology, 440 UCB, Boulder, CO 80309-0440, USA}

\author[0009-0000-9963-6874]{B. Molina}
\email{brian.leftraru@gmail.com}   
\affiliation{CePIA, Astronomy Department, Universidad de Concepci\'on,  Casilla 160-C, Concepci\'on, Chile}

\author[0000-0003-1315-3412]{M. L. Lister}
\email{mlister@purdue.edu}   
\affiliation{Department of Physics and Astronomy, Purdue University, 525 Northwestern Avenue, West Lafayette, IN 47907, USA}

\author[0000-0002-0491-1210]{E.R. Most}
\email{emost@caltech.edu}
\affiliation{TAPIR, Mailcode 350-17, California Institute of Technology, Pasadena, CA 91125, USA}
\affiliation{Walter Burke Institute for Theoretical Physics, California Institute of Technology, Pasadena, CA 91125, USA}

\author[0000-0003-2454-6828]{S. O'Neill}
\email{so3803@princeton.edu}
\affiliation{Owens Valley Radio Observatory, California Institute of Technology,  Pasadena, CA 91125, USA}
\affiliation{Department of Physics, Princeton University, Princeton, NJ 08544, USA}


\author[0009-0005-7899-3705]{V. A. Diamantopoulos}
\email{vasdiam@physics.uoc.gr} 
\affiliation{Department of Physics, University of Crete, GR-71003 Heraklion, Greece}
\affiliation{Institute of Astrophysics, Foundation for Research and Technology-Hellas, GR-70013 Heraklion, Greece}

\author[0000-0001-9011-0737]{N. Globus}
\email{globus@astro.unam.mx}
\affiliation{Instituto de Astronomía, Universidad Nacional Autónoma de México Campus Ensenada, Carr. Tijuana-Ensenada km107, Ensenada, BC22800, Mexico}

\author[0000-0002-2024-8199]{T. Hovatta}
\email{talvikki.hovatta@utu.fi}   
\affiliation{Finnish Centre for Astronomy with ESO (FINCA), University of Turku, FI-20014 University of Turku, Finland}
\affiliation{Aalto University Department of Electronics and Nanoengineering, PL~15500, FI-00076 Espoo, Finland}
\affiliation{Aalto University Mets\"ahovi Radio Observatory,  Mets\"ahovintie 114, 02540 Kylm\"al\"a, Finland}

\author[0000-0001-9200-4006]{I. Liodakis}
\email{yannis.liodakis@gmail.com}   
\affiliation{Institute of Astrophysics, Foundation for Research and Technology-Hellas, GR-70013 Heraklion, Greece}

\author[0000-0001-5213-6231]{T. J. Pearson}
\email{tjp@astro.caltech.edu}   
\affiliation{Owens Valley Radio Observatory, California Institute of Technology,  Pasadena, CA 91125, USA}

\author[0000-0001-7662-2576]{S. E. Tremblay}
\email{strembla@nrao.edu}   
\affiliation{National Radio Astronomy Observatory, 1011 Lopez Road, Socorro, NM 87801, USA}

\begin{abstract}
We present evidence of 2:1 harmonic periodicities in two supermassive black hole binary (SMBHB) candidates (PKS J0805-0111 and PKS 2131-021) in the Owens Valley Radio Observatory (OVRO) + University of Michigan Radio Astronomical Observatory (UMRAO) monitoring programs. This potential discovery follows the previous detection of a 2:1 harmonic periodicity in the SMBHB candidate J1309+1154 from these programs. If confirmed, these suggest a physical origin of blazars exhibiting strong periodicities favoring 2:1 commensurabilities in their light curves.  There are significant differences between the 2:1 harmonics in these three blazars, which may be due to differences in their central engines. These three SMBHB candidates demonstrate the importance of careful dissection, i.e. cutting  apart and close examination, of blazar light curves, to reveal underlying evidence of coherence between different aspects of the periodicities observed in blazars in the presence of correlated random noise variations.  This is particularly important for blazar SMBHB searches, given that the observed periods thus far are all longer than 2 years, and can disappear for long timespans.  Harmonics may therefore be crucial in confirming the reality of periodicities and identifying blazar SMBHB candidates. We discuss four models that can account for 2:1 commensurable periodicities in blazar light curves and explain how future detections of shorter period SMBHB candidates, or the lack thereof, may be used to discriminate among the models.   Arguments for and against the reality of these periodicities are presented.

\end{abstract}


\section{Introduction}\label{sec:intro}

Blazar light curves are generally dominated by large random flares.  However, in the Owens Valley Radio Observatory (OVRO) 40 m Telescope monitoring program \citep{2011ApJS..194...29R}, two blazars with sinusoidal variations dominating the light curves have been found: PKS J0805$-$0111 and PKS 2131$-$021   
(\citealt{2021MNRAS.506.3791R};
\citealt{2021RAA....21...75R};
\citealt{2022ApJ...926L..35O} [Paper I];
\citealt{2025ApJ...985...59K} [Paper II];
\citealt{2025ApJ...987..191D} [Paper III]. The combination of the OVRO and University of Michigan Radio Astronomical Observatory  (UMRAO) monitoring programs has recently revealed a third such blazar, PKS J1309+1154 (\citealt{2026ApJ..1000...72M} [Paper IV];
\citealt{2026ApJ...996L..39R} [Paper V]).

 In Paper IV it is shown that  the steep slope of the power spectral density (PSD) spectrum of blazars,   coupled with the random nature of blazar flares, produces light curves whose variability power spectra, as revealed by generalized Lomb-Scargle (GLS) periodograms
\citep{1976Ap&SS..39..447L,1982ApJ...263..835S,2009AandA...496..577Z}, and weighted wavelet-Z (WWZ) transforms \citep{ragazzini1952analysis,1996AJ....112.1709F}, 
 are dominated by  randomly-generated spurious periodicity peaks and  harmonics, that  have nothing to do with orbital motion or precession of the supermassive black holes (SMBHs) producing the jets. Therefore, differentiating between genuine periodicities generated by the central engines in AGN, and random periodicities generated by the steep PSD spectrum and the stochastic nature of blazar flares, is challenging \citep{1978ComAp...7..103P,2016MNRAS.461.3145V}.

 

The discovery of a coherent 2:1 harmonic in PKS J1309+1154 (Paper V) has led us to formulate the hypothesis that strong supermassive black hole binary (SMBHB) candidates may commonly exhibit 2:1 commensurate periodicities. The analyses presented in this paper were therefore carried out as a test of this independently motivated hypothesis by searching the previously studied light curves of PKS 2131-021 and PKS J0805-0111 for evidence of such commensurabilities.

Here we discuss SMBHB candidates identified through their radio light curves.  Readers interested in SMBHB candidates identified in other wavebands are referred to the recent review by \citet{Dorazio2023}, and papers on SMBHB candidates identified through their optical or Gamma-ray emission \citep[see, e.g.,][]{Sillanpa1988,LehtoVal1996,2015ApJ...813L..41A,Graham2015,2016ApJ...819L..37V,2021MNRAS.503.4400D}.

While one might reasonably expect to see periodicities related to the dynamics of the central engines in the light curves of SMBHB candidates, there was, until recently, no obvious reason why in some cases the variations might be sinusoidal in nature \citep[see, e.g.,][]{2026ApJ...997..316L}. However, the results from the OVRO and UMRAO monitoring programs show that sinusoidal variations do dominate the light curves of $0.1\%-1\%$ of blazars (Papers I, II and IV). This led us to explore the effects of orbital motion on blazar emission, which is strongly beamed towards the observer.  This orbital motion does indeed produce sinusoidal variations in flux density, as was first revealed in the  model of \citet{2017MNRAS.465..161S}, and is described in detail in Papers I and II, and \citet[hereafter Paper VI]{2026ApJ...997...85S}.  

Motivated by the observed sinusoidal periodicity in PKS 2131$-$021, an alternative search method for periodicities and harmonics in blazar light curves was introduced in Paper I --- detailed analysis of the light curves using a combination of careful visual inspection, coupled with sine-wave fitting and subtraction, closely following the method used so successfully in radio astronomical image reconstruction \citep[see, e.g.,][]{1984ARA&A..22...97P}.   We call this the Sine Wave Fitting, or SWF, method --- see Paper IV for a step by step description of this approach to analyzing blazar light curves.

\begin{figure*}[!t]
   \centering
   \includegraphics[width=0.9\linewidth]{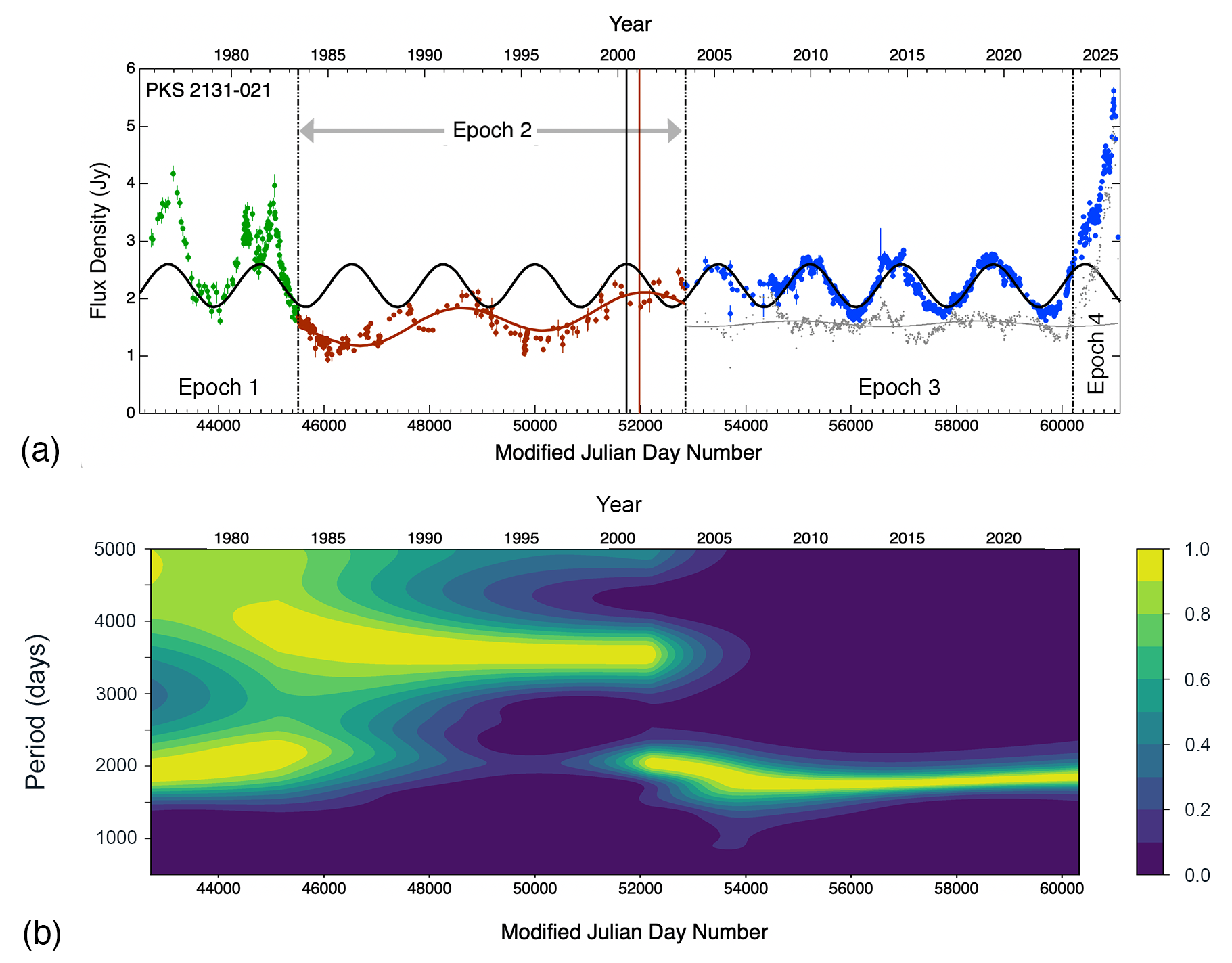}
   \caption{The PKS 2131$-$021 commensurate periodicities. (a) The Epoch 1 (green symbols), Epoch 2 (brown symbols), and Epochs 3 and 4 (blue symbols) light curves from 1975 to 2026. The black sinusoid is the fit to the Haystack+OVRO (Epoch 1 + Epoch 3) data, taking  correlated variations into account, from Paper II. The brown trended sinusoid is the fit to the  data in Epoch 2 taking account of correlated variations (see Appendix \ref{sec:corrnoise}).  The solid black vertical line marks a maximum of the black sinusoid. The brown vertical line marks the corresponding maximum of the brown sinusoid. The difference between these two maxima  is 245.9 days, i.e., 7.1\% of the Epoch 2 periodicity. In Epochs 3 and 4 the gray symbols show the residual after subtraction of the sinusoidal component (black curve). The gray sinusoid is the least squares fit of the subharmonic to the residual in Epoch 3, which has amplitude 47 mJy and reduces the rms variation by only 2.7\%  (see text). (b) Time-normalized WWZ plot of the light curve of PKS 2131$-$021 in Epochs 1, 2 and 3.   We  normalized each 100-day time bin by dividing the  power across all the periods in each bin by the maximum power in that bin. Thus every time bin has a maximum power of unity.  This enables us to see how the periods change over the span of the observations, and shows clearly the factor 2:1 commensurability of the sinusoidal periodicities between Epoch 2 and Epochs 1+3.  Note that the transition from the longer to the shorter period in the WWZ plot occurs $\sim 1800$ days before the start of Epoch 3 (see text).}
   \label{plt:PKS2131harmonic}
\end{figure*}

We show in this paper that there are strong harmonic 2:1 commensurabilities  in both PKS J0805$-$0111 and PKS 2131$-$021, but that, unlike the case in PKS J1309$-$1154, these are either not (PKS J0805$-$0111) or may not be (PKS 2131$-$021) in phase with the fundamental.   The absence of a clear phase relationship relative to the fundamental casts some doubt on the authenticity of the subharmonics we find evidence for in these two sources. However, the fact that there is evidence of strong 2:1 commensurabilities in all three strong SMBHB candidates found thus far in the OVRO+UMRAO monitoring programs is clearly worth investigating because of its possible relevance to searches for SMBHB candidates, as well as  the light it  might cast  on the central engines of blazars.  

We explore the evidence for periodicities and harmonics/subharmonics in PKS J0805$-$0111, PKS J1309+1154, and PKS 2131$-$021 in terms of both the modified kinetic orbital (MKO) model of Paper VI in which the emission regions are within a few pc of the central engine, and in terms of a new model: a binary consisting of a single SMBH and a much less massive ($\sim 100 M_\odot$) secondary, such as a Wolf--Rayet star.  This last model can accommodate all of the phenomenology observed thus far in these three objects.

It has not escaped our attention that, in this scenario, any massive star on a trajectory that would carry it close to the SMBH is likely to lead to a Tidal Disruption Event (TDE), which might possibly evolve into a Compact Symmetric Object (CSO) \citep{1994cers.conf...17R,2024ApJ...961..240K,2024ApJ...961..241K,2024ApJ...961..242R,2024ApJ...977..195D, 2024MNRAS.528.6302S, 2025ApJ...987...26S}.  Obvious issues to be addressed in future are whether there is a natural way to form or capture such massive stars in blazar disks and how to get them to the tidal radius before they explode or collapse.

\vskip 6pt
\noindent
\textit{Terminology:} In this paper we call the stronger periodicity the ``fundamental'' and the weaker periodicity the ``harmonic'' if its frequency is twice that of the fundamental, and the ``subharmonic'' if its frequency is half that of the fundamental. No particular physical significance is attached to these terms --- they are used here merely as convenient labels.

\section {Coherence in the Light Curves of SMBHB Candidates}\label{sec:coherence}

The observation of periodic signals in the light curves of some blazars is evidence of coherent  signals with a predictable phase behavior, and  this enables us to select likely SMBHB candidates.  The existence of these periodic signals demonstrates that not all blazar light curves are entirely stochastic in nature. 

Coherence in blazar light curves is most easily, and most frequently, identified through  a clear, dominant periodicity.  A good example is that of PKS 2131$-$021 (Papers I, II). In this particular case, coherent sinusoidal variations were seen from 1975 to 1983, then disappeared for 19 years, and then returned with the same period and in phase, in 2003, and persisted until 2025.  An additional type of coherence has been reported in PKS J1309+1154 (Papers IV, V) in which both a fundamental periodicity and a first harmonic periodicity (i.e., a periodicity of twice the frequency of the fundamental) have been observed in phase with the fundamental.  In this case, the light curve is seen to oscillate about the fundamental twice per period throughout the light curve, showing that the harmonic is coherent with the fundamental throughout the light curve.  

These two examples of coherence in SMBHB candidate light curves are strikingly different, and each of them was discovered not by automated computer analysis of a large number of light curves, but by careful visual inspection of the light curves themselves.  Other types of (as yet unknown)  coherence in SMBHB candidate light curves should therefore be anticipated and searched for.  We present two such examples of previously unknown and unexpected coherence in this paper.

\section{Evidence for a Subharmonic in PKS 2131$-$021}\label{sec:2131}

The light curves of PKS 2131$-$021 from the UMRAO, the OVRO, and the Haystack Observatory, at 14.5/15/15.5 GHz , respectively,  have been discussed extensively in Papers I and II. We present the 50-year light curve from these three observatories in Fig.  \ref{plt:PKS2131harmonic}(a), where, as in Papers I and II, the light curve has been divided into Epoch 1 (MJD$<45500$), Epoch 2 ($45500<$ MJD $<52850$), and Epoch 3 (MJD $>52850$).   

In view of the large flare, which is significantly larger than any of the previously observed flares in this object, we now define a new epoch  in  the light curve of PKS 2131-021: Epoch 4 (MJD $\geq$ 60095). Thus the third epoch is now re-defined as $52850<$MJD$<60095$. 

In Fig.~\ref{plt:PKS2131harmonic}(a) the Paper II sinusoidal fit to the data in Epochs 1 and 3 is shown by the black curve, which has period $P=1739.8\pm17.4$ days. In this fit, the correlated errors have been taken into account, as described in Appendix B of Paper II.  In Paper I the kinetic orbital  (KO) model was introduced to account for the periodicity seen in Epochs 1 and 3, and it was pointed out that harmonics are expected in this model.  We return to this point in \S \ref{sec:modeling}. In Paper II a detailed  analysis was carried out, to no avail, in search of harmonics in the light curves of Epochs 1 and 3, and upper limits were placed on the amplitude of any such harmonics.  

In \S 3.4 and Table 1 of Paper I, it was pointed out that there is a 2:1 commensurability in PKS 2131$-$021 between the coherent periodicity of Epochs 1+3 and an apparent periodicity  in Epoch 2.  As stated in Paper I,  this was not followed up at the time because of the absence of other evidence that this might be a real commensurability.
This situation changed after the discovery of a coherent 2:1 harmonic in PKS J1309+1154 (Paper V). That result motivated the hypothesis of this paper that 2:1 commensurabilities may be a generic property of at least some SMBHB candidates. We therefore returned to the previously published PKS 2131--021 light curve to test this hypothesis in Epoch 2 of the light curve of PKS J2131$-$021, having already ruled out any such in Epochs 1 and 3 (Paper II).

In Fig.~\ref{plt:PKS2131harmonic}(b) we present the normalized WWZ transform of the light curve, first presented in Paper I.   The figure shows that the 2:1 transition from the long to the short period towards the end of Epoch 2 occurs $\sim 1800$ days before the transition to Epoch 3, but  it can be seen from Fig.~\ref{plt:PKS2131harmonic}(a) that the subharmonic dies out about 1000 days before the start of Epoch 3, and that there is a narrow, out of phase, peak around MJD 51400.  These two features of the light curve explain this apparent early switchover in the dominant periodicity  in the WWZ spectrum. It was decided, in Paper I, to set the transition from Epoch 2 to Epoch 3 at MJD=52850 because this is where the sinusoidal emission of Epoch 3 first becomes clearly visible.

In Paper I we carried out an analysis of the Epoch 2 light curve and found a strong peak in the GLS spectrum at $P\sim 3900$ days.  We then carried out an SWF anlysis, with a uniform prior for the period, which returned a period of $P\sim 3779 \pm 46$ days (see Table \ref{tab:bestfit} of Paper I). 

We re-analyzed the Epoch 2 light curve to test our hypothesis by repeating the analysis carried out in Paper I, but replacing the uniform prior with a normally-distributed prior of $P=3479.6+/-34.8$ days, based on the period derived for the Epoch 1+3 light curves  (Paper II), and this revealed a sinusoidal periodicity of amplitude 0.26 Jy, shown by the brown curve in Fig.~\ref{plt:PKS2131harmonic}(a).  The parameters of these sinusoidal fits are shown in Table \ref{tab:bestfit}. Note that a linear trend was subtracted from the original light curve before carrying out these fits.

\begin{table*}[!t]
\centering
\caption{PKS 2131$-$021 sine wave fits to Epoch 2 light curve and phase comparison to Epochs 1 and 3}
\begin{tabular}{lccc}
\hline \hline
Parameter&Uniform Prior&Normally Distributed Prior$^*$\\
&(Paper I)&$P=3479.6 \pm 34.8$ days\\
\hline
$P$ (days)         &$3779.1 \pm 46.0$& $3481 \pm34$ \\
$\phi_0$ (rad)     &$0.35^\dagger \pm 0.08$& $4.35^\ddagger \pm 0.63$ \\
$A$ (Jy)           &$0.39  \pm 0.02$& $0.26 \pm 0.10$ \\
$S_0$ (Jy)         &$1.72 \pm 0.02$& $2.24 \pm 0.07$ \\
log$_{10}S_1$&none &$-4.11^\ddagger \pm 0.04$ \\
Phase difference/Epoch 2 Period &0.6\%&7.1\%\\
\hline
\end{tabular}
\tablecomments{ $^*$This sine wave fit includes the effect of correlated errors (see Appendix \ref{sec:corrnoise}). $^\ddagger$Indexed to MJD 57000.  The Paper I sine wave does not include the effects of correlated errors. $^\dagger$Indexed to MJD 51000.  The facts that both the phase alignment of the Epochs 1+3 sine wave to within 0.6\% of the fit with uniform prior in Paper I, which does not take correlated noise into account, and the period agreement ($3481\pm34$ days \textit{vs.\/} $3479.6\pm 34.8$ days) are so close, appear to be coincidental, and cannot be used in calculating the significance.}
\label{tab:bestfit}
\end{table*}

\begin{table}[!t]
\centering
\caption{PKS 2131$-$021 Significance Tests}
\begin{tabular}{cccc}
\hline \hline
Test& Simulations that& p-value & $\sigma$\\
&passed the test&&\\
\hline
\#1 Period&8/10000 & $8\times 10^{-4}$&3.2\\
\#2 Phase&-&0.14&1.1\\
\hline
\end{tabular}
\tablecomments{Test \#1 is  the number of simulations for Epoch~2  in the $\pm 3\sigma$ wide  period window  P=3375.2-3584.0 days  that have  p-value $<$ the observed p-value (see Paper I). }
\label{tab:significance}
\end{table}

The period we derive in this way is $3481\pm 34$ days, which may be compared to twice the period derived for epochs 1 and 3 combined in Paper II of $3479.6\pm 34.8$ days, i.e., a difference of $1.4\pm 49.7$ days.  The corresponding maxima of the Epoch 1+3 sinusoid and the Epoch 2 sinusoid are indicated by the black and brown vertical lines in Fig.~\ref{plt:PKS2131harmonic}(a), which occur at MJD 51733.4 and MJD 51979.3, respectively.  Thus these maxima differ by 245.9 days, which is 7.1\% of the Epoch 2 period. 

Paper I presented the results for  the \textit{global\/} significance of the periodicity in Epoch 2 (p-value=$6.32\times 10^{-2}$), i.e., the significance in the case where the period is completely unknown.   Under our present hypothesis of  2:1 commensurabilities in the periods found in SMBHB candidates, and given that this is based on PKS J1309+1154, when applied to PKS 2131--021 this amounts to \textit{a priori\/} knowledge of the period we are looking for in Epoch 2, namely that it is either half or twice the period in Epochs 1+3.  From Paper II, and taking the $\pm 3 \sigma$ values for the uncertainty, the range of periods to be considered is $2 \times 1739.8\pm 52.2$ days, i.e., 3375.2--3584.0 days.

We have carried out simulation tests on the light curve of PKS 2131$-$021 in Epoch 2,  with simulations generated as described in Papers I and II, in which both the PSD and the probability density function (PDF) are taken into account. We use the PSD slope of the Epoch 2 light curve of $\beta = -1.75\pm0.26$ found in Paper I. The results are shown in Table \ref{tab:significance}.  From 10,000 simulations of the Epoch 2 light curve, we find that only 8 simulations in the period window $P=3375.2-3584.0$ days have p-values less than the observed p-value (Paper I).  Thus the probability of an observation that is more extreme is given by p-value=$8 \times 10^{-4}$ (i.e. $3.2 \sigma$).  

\textit{According to the above test, therefore, it is  highly unlikely that the Epoch 2 period would be this close to twice that of the Epoch 1+3 period by chance, and we henceforth regard it as real.}

In  Table \ref{tab:significance}, Test \#2 is the phase test given by the ratio of the  MJD difference in phase between the black and brown curves of Fig.~\ref{plt:PKS2131harmonic}, which is $7.1\%$ of the period in Epoch 2, so allowing for $\pm 7.1\%$ gives a p-value of twice this, i.e., there is a 1 in 7 random chance of the phases lining up as well as they do. A 1 in 7 chance is not low, so this does not amount to significant evidence that the phases between the Epoch 2 and the Epoch 1+3 periodicities are aligned, i.e., they may well not be aligned.

\begin{figure*}[!t]
   \centering
   \includegraphics[width=1.0\linewidth]{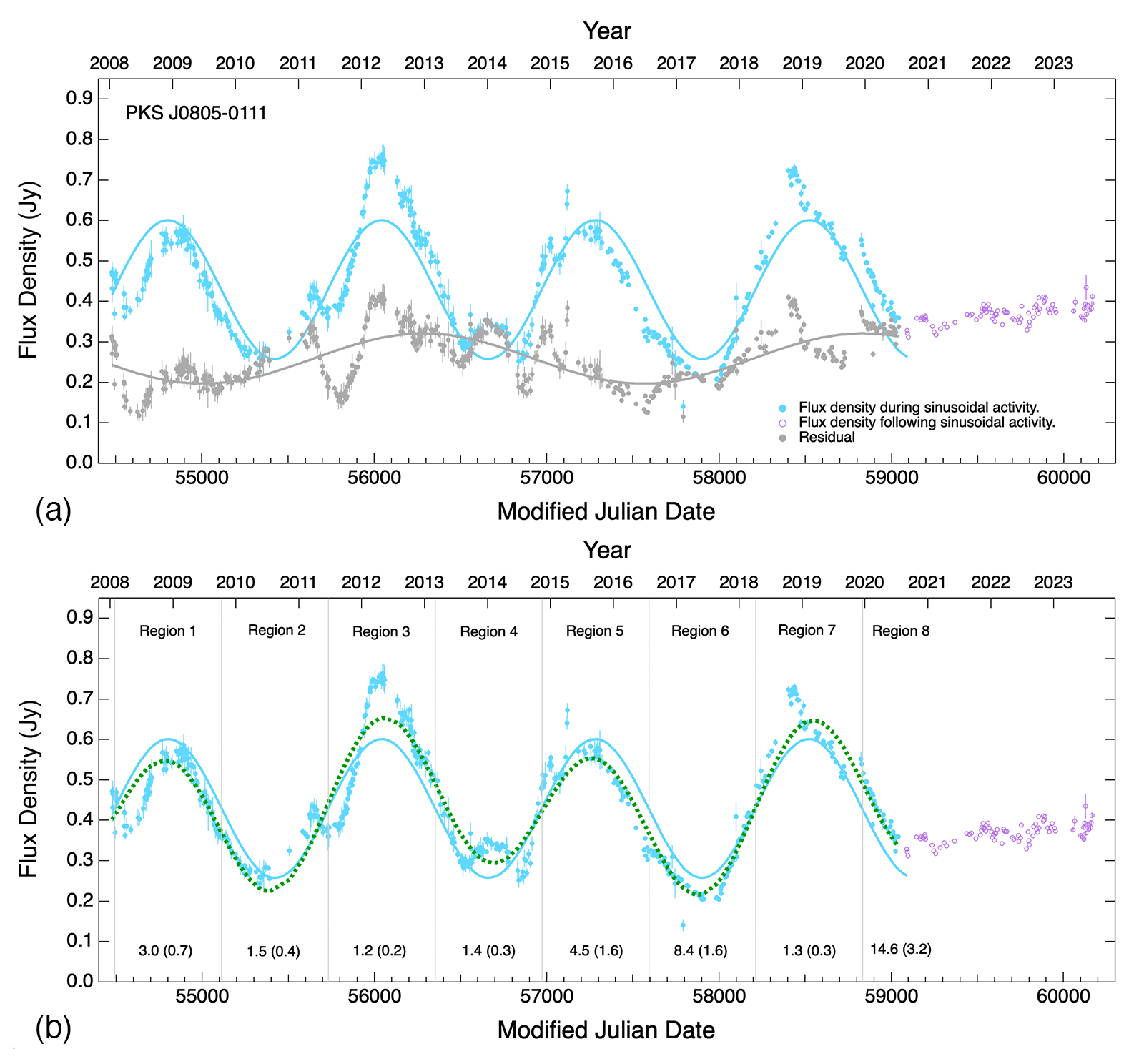}
   \caption{The PKS J0805$-$0111 light curve. The blue data points show the OVRO light curve during the period up to MJD 59041 when the periodicities were present.  The purple data points show the OVRO light curve after the periodicities ceased. (a) the light blue curve is the fit to the blue data points --- this is the fundamental periodicity. The gray points show the residual light curve after subtracting the fundamental. The gray curve shows the sinusoidal fit to the residual data --- this is the subharmonic periodicity, which has $P=2550.0\pm 43.8$ days, showing a 2:1 commensurability to within 2.8\%, with respect to the fundamental (see text). (b) As in (a) but now with the green dotted curve showing the fit when the subharmonic is added to the fundamental. The improvement of the fit when the subharmonic is added is clear. The vertical light gray lines divide the light curve into half-period regions of the fundamental. In each region the ratio of the rms deviation from the blue curve to that from the green dot-dashed curve is shown, with the uncertainty in the ratio given in brackets. In all 8 of the 8 regions (p-value = $3.9\times 10^{-3}$ (2.7$\sigma$)) the ratio is greater than 1, showing that the addition of the subharmonic gives a better, and hence more  coherent,  fit to the data.  The ratio of the rms deviation from these two curves for the eight regions combined is   $2.62 \pm 0.27$, i.e., a $6\sigma$ difference from unity (see text).}
   \label{plt:PKSJ0805harmonic}
\end{figure*}


\section{Evidence for a Subharmonic in PKS J0805$-$0111}\label{sec:0805}

The OVRO 15 GHz light curve of PKS J0805$-$0111 was presented in Paper III. Unfortunately this object was not observed at UMRAO.  The OVRO light curve from Paper III, plus more recent data from 2023 is shown in Fig.~\ref{plt:PKSJ0805harmonic}.  The light blue sinusoid is the least squares sine wave fit to the observed light curve up to MJD 59041 taken from Paper III.  The corresponding values of the parameters are given in Table \ref{tab:J0805}.

The residual light curve, after subtraction of the blue sine wave, is shown by the gray points in Fig.~\ref{plt:PKSJ0805harmonic} and was included in Paper III. In that paper (Appendix C), the residual light curve  was discussed in some detail, and it was shown that there is a sine wave, of period 2445 days, for which the global p-value=0.18. While this is of the right frequency to be a subharmonic of the dominant sine wave, this was not pursued in Paper III because there did not seem to be any reason for doing this.  However, given the 2:1 first harmonic in PKS J1309+1154, we decided to carry out the SWF analysis on the residual light curve of PKS J0805$-$0111 shown by the gray sinusoidal curve in Fig.~\ref{plt:PKSJ0805harmonic}. The parameters of the fit are given in Table \ref{tab:J0805}.  It can be seen there that the freely-fitted sine wave to the residual has a period of $2550.0\pm 43.8$ days, which may be compared to the 2:1 subharmonic of the sine wave of the observed light curve, i.e., to $P=2481.4 \pm 9$ days. These two periods differ in ratio by 2.8\%, and periods by  $68.6\pm 44.7$ days. 

\begin{table}
\caption{PKS J0805$-$0111 best-fit sine-wave parameters}
\label{tab:J0805}
\centering
\footnotesize
\begin{tabular}{lcc}
\hline \hline
Parameter&Observed light curve&Residual light curve\\
\hline
$P$ (d)       & $1240.7 \pm 4.5$      & $2550.0 \pm 43.8$ \\
$A$ (Jy)      & $0.1709 \pm 0.0043$   & $0.0621 \pm 0.0028$\\
$\phi$ (rad)  & $-0.142 \pm 0.024$    & $2.950 \pm 0.053$\\
$S_0$ (Jy)      & $0.4294 \pm 0.0030$   & $0.009 \pm 0.0021$\\
$\sigma_0$      & $0.0662 \pm 0.0022$   & $0.0466 \pm 0.0016$\\
\hline
\end{tabular}
\tablecomments{In Appendix \ref{sec:sinefits} we present a two-component sinusoidal model fit that returns very similar values to those given in this section. }
\end{table}

\begin{table*}
\centering
\caption{Reduction factors when subtracting Fundamental and Harmonic: Observed \textit{vs.\/} Simulated Light Curves}
\begin{tabular}{lcccccccccc}
\hline \hline
Periodicity	&	PKS	&	PKS	&PKS&PKS&	Sim \#	&	Sim \#	&	Sim \#	&	Sim \#	&	Sim \#	&	Sim \#	\\
	&	J0805&	J1309		& 2131&2131&	79701	&	97484	&	157884	&	464731	&	589833	&	765735	\\
    &&&Epoch 2&Epoch 3&&&&&&\\
\hline
Fundamental$^{\textrm{f}}$	&	50\%$^{\textrm{f}}$&	65\%	$^{\textrm{f}}$	&-&53\%$^{\textrm{f}}$&	22\%	&	61\%	&	42\%	&	17\%	&	43\%	&	60\%	\\
Harmonic$^{\textrm{h}}$/Subharmonic$^{\textrm{s}}$	&	27\%$^{\textrm{s}}$&	20\%$^{\textrm{h}}$	&44\%$^{\textrm{s}}$	&2.7\%$^{{\textrm{s}}\dagger}$&	8\%	&	1\%	&	4\%	&	5\%	&	6\%	&	3\%	\\
Combined	&	63\%&	72\%	&-	&53\%&	28\%	&	61\%	&	45\%	&	21\%	&	46\%	&	62\%	\\
\hline
\end{tabular}
\tablecomments{Values for PKS J0805$-$0111 are up to MJD 59041. $^\dagger$Note that the ``subharmonic'' in PKS 2131$-$021 Epoch 3 is \textit{not\/} a believable periodicity, showing that in this source the fundamental and subharmonic really do alternate (see text).}
\label{tab:J0805rms}
\end{table*}

In   Appendix \ref{sec:sinefits} we present the results of simultaneous fitting of two sinusoidal curves.  We see there that the freely fitted least-squares sine wave to the residual has a period of $2541.6 \pm 36.1$ days. This is  only 4\% smaller than twice the period of the single sine sine wave fit to the original data  (Model \#1 in Table \ref{tab:J0805}). Four models are presented in Appendix \ref{sec:sinefits}. Model \#1  consists of a single sine wave. In Models \#2, \#3, and \#4 two sine waves have been fitted simultaneously.  In Model \#2 the periods have ratio 2:1 and the phases are constrained to be the same.  As can be seen, the periods are $1240.8 \pm 4.5$ days and twice this, and the amplitude of the subharmonic is $<0.8$ mJy.  So there is strong evidence against a subharmonic that is in phase with the harmonic. In Model \#3 the periods again have a 2:1 ratio but the phases are independent.  In Model \#4 both the periods and the phases are independent. The ratio of the subharmonic to the fundamental   period is $2.08\pm0.03$.  The Bayesian Information Criterion (BIC) scores for models \#3 and \#4 differ by only 2.5.  In other words, there is no strong evidence that the period $P_2$ differs from  twice the period $P_1$.

\begin{figure}[t]
   \centering
   \includegraphics[width=1\linewidth]{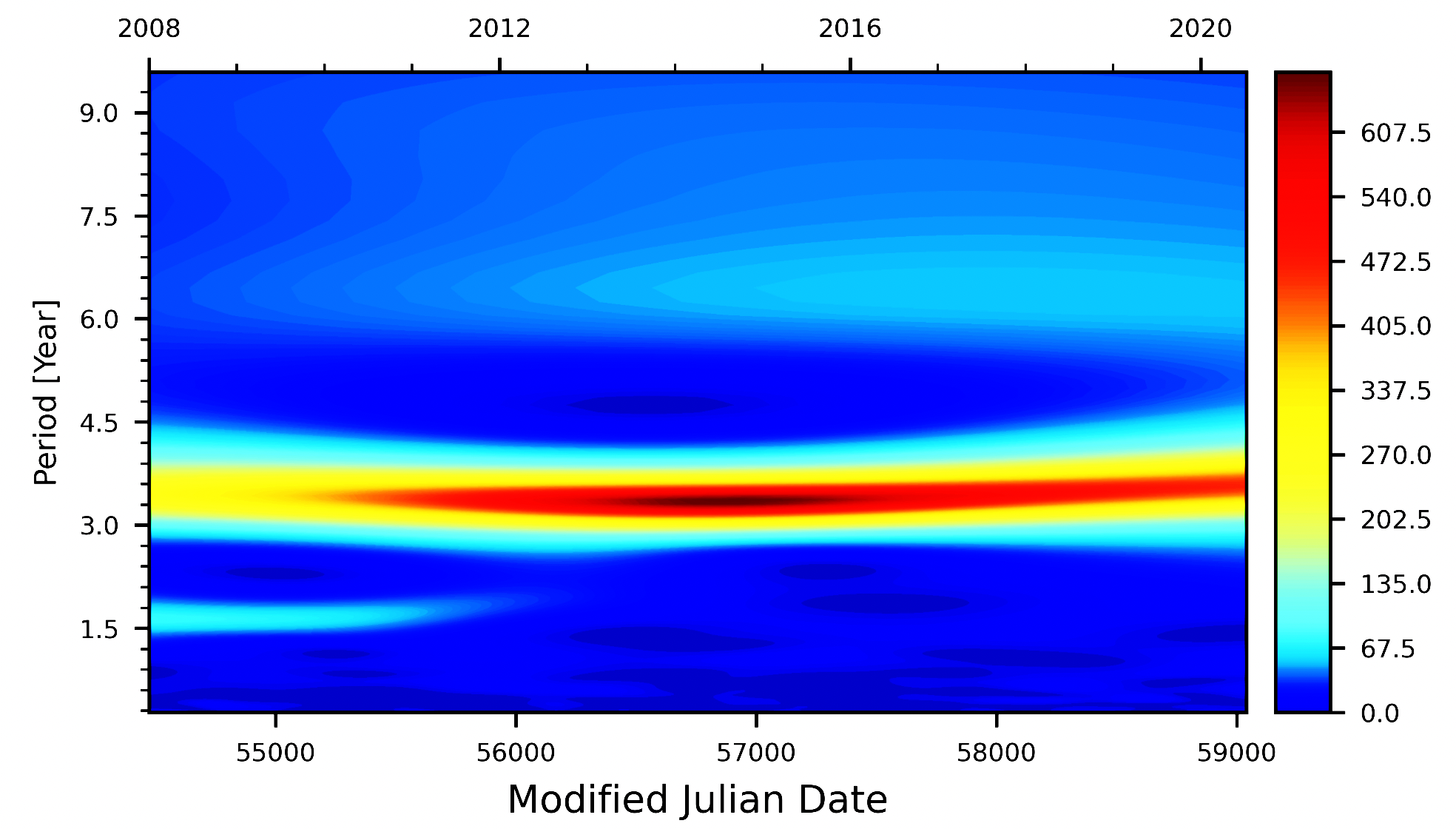}
   \caption{The OVRO WWZ spectrum of PKS J0805$-$0111 up to MJD 59041.  This is the spectrum using the square root of the power (see text).  Note that both the subharmonic ($\sim$7 yr) and the fundamental ($\sim$3.4 yr) periodicities are visible over the full light curve, showing coherence over this 12.5-year period.
   }
   \label{plt:J0805WWZ}
\end{figure}

The WWZ spectrum of PKS J0805$-$0111 is shown in Fig.~\ref{plt:J0805WWZ}, where we have used the square root of the power (see Paper V) in order to bring out the subharmonic more clearly.  As in the case of PKS J1309+1154, the two periodicities are clearly seen over the whole length of this light curve, until  both ceased after MJD 59041 (7/11/2020).

\subsection{The Search for  Harmonics in PKS J0805$-$0111}\label{sec:searching}

It is interesting to note the difference between the approach of the independently fitted sinusoids to the observed and residual light curves, and the jointly fitted sinusoids to the observed light curve.  While these are very similar, they are not mathematically identical, because in fitting the first sine wave alone there are fewer variables and more emphasis is placed on trying to determine whether or not there is a real periodicity.  As pointed out in Paper IV, the procedure of successively subtracting the fitted sine waves is similar to the common practice of subtracting successive point spread functions in aperture synthesis  image formation \citep{1984ARA&A..22...97P}.  The great advantage of this approach is that the residual light curve is carefully examined after each component is subtracted, in order to ensure that there is clear evidence of a sinusoid in the residual curve.  That there clearly is such evidence in the cases of both PKS J1309+1154 and PKS J0805$-$0111 can be seen in Fig.~1(b) of Paper V and Fig.~\ref{plt:PKSJ0805harmonic} of this paper, respectively.

In Fig.~\ref{plt:PKSJ0805harmonic}(b) it is clear that the quality of the fit when the subharmonic is included is much better than the fit without the subharmonic.  The combination of all the data up to MJD 59041 yields a ratio of the rms deviations without the subharmonic to the rms deviations with the harmonic of $2.62 \pm 0.27$, i.e., a $6\sigma$ difference from unity. But it should be kept in mind that this asssumes that there are no long-term noise variations, which is clearly not the case in view of the steep slope of the PSD. Nevertheless, the improvement in the fit when the subharmonic is taken into account is very clear to see in Fig.~\ref{plt:PKSJ0805harmonic}(b).

In order to test the coherence between the subharmonic and the fundamental over the whole 12.5 years for which the periodicity is seen, we have divided the light curve into 8 regions of length equal to half that of the fundamental and calculated the ratio of the rms deviations of the fits with and without the subharmonic. The ratios of the rms deviations are given at the bottom  of each region, together within their uncertainties (in brackets).  These show that the fit with the subharmonic is better than the fit without it  in all eight regions, and that in four of the eight regions the ratio exceeds unity by more than $2\sigma$, and in three of the eight regions by more than $3\sigma$. 

\textit{The total ratio of the rms deviations, and the individual ratios of the  rms deviations in the eight half-period regions, taken  together, demonstrate decisively that the fit with the subharmonic is far better than the fit without it. For this reason, we regard the subharmonic in PKS J0805$-$0111 to be real, and not just a random occurrence.}

The phases of the subharmonic and the fundamental in Model \#3 and Model \#4 differ by 270.3 and 362.3 days, respectively.  The mean is 316.3 days, which is an offset in phase of 25.9\% (i.e., $\sim \pi/2$) relative to the period of the fundamental. This offset is much larger than the uncertainties, and it might be significant in terms of finding models that explain this behavior.

\section{The Strengths of the 2:1 Commensurabilities in the 3 SMBHB Candidate Light Curves}\label{sec:commens}

It is important to note that the fundamental and harmonic/subharmonic periodicities in the three blazars in the OVRO+UMRAO samples are all \textit{strong periodicities\/} in the sense that they contribute a significant fraction of the variability in these three blazars.  In Table \ref{tab:J0805rms} we show the reduction factors in the rms variations in these light curves when the fundamental and the harmonic/subharmonic periodicities are subtracted. As can be seen here, these are fairly similar for the three sources, with the subtraction of the fundamental reducing the rms variations by 50\%-65\%, and the subtraction of the harmonic/subharmonic reducing the rms variation by 20\%-44\%.   In the case of PKS 2131$-$021 Epoch 3, a subharmonic was fitted, but it has very small amplitude, and the reduction factor of only 2.7\% demonstrates that it is not a periodicity that can be trusted.  Likewise, in Paper II, a harmonic of the fundamental was searched for, and found to be very small and hence not  a periodicity that can be trusted.

 Also shown in  Table \ref{tab:J0805rms}  are the same reduction factors, for 2:1 harmonics in six randomly chosen simulations, out of the $10^6$ simulations of the light curve of PKS J0801-0111 of Paper III.  As can be seen clearly here, the reduction factors for the harmonics are all in the range 1\%-8\% and hence significantly smaller than are seen in the cases of PKS J0805$-$0111, PKS J1309+1154 and PKS 2131$-$021.  In addition to the rms reduction factors being significantly larger in the real light curves, there is also the fact that the harmonics coherently improve the fit over the whole light curve, where we see them simultaneously in PKS J0805$-$0111 and PKS J1309+1154, as shown in Paper V and in Fig.~\ref{plt:PKSJ0805harmonic}(b), respectively. 

 Thus, of $\sim 1830$ blazars monitored on the OVRO 40 m Telescope at 15 GHz from 2008 to 2025, only three, PKS J00805$-$0111, PKS J1309+1154 and PKS 2131$-$021, show strong sinusoudal fluctuations over a significant portion of their light curves; and \textit{all three\/} of these show  strong evidence of 2:1 commensurabilities in their light curves, which, if confirmed, suggests that blazars exhibiting strong periodicities  also tend to show 2:1 commensurabilities in their light curves.

\section{The Incidence of SMBHB Candidates in the OVRO+UMRAO Samples}\label{sec:incidence}

In this section we compare the numbers of SMBHB candidates found in the 16-year OVRO light curves with those found in the 46-yr UMRAO+OVRO light curves. A more detailed discussion of this may be found in Paper IV. We summarize the main points briefly here. We then discuss the fact that only SMBHB candidates with rest frame periods $P>1$ year have been found thus far, and the absence of SMBHB candidates with periods shorter than one year.
\vskip 6pt
\noindent
\textit{A Successful Prediction:} The argument below was first presented as a prediction in a grant proposal in November 2023, six months before an analysis of the combined OVRO+UMRAO sample identified PKS J1309+1154 as an SMBHB candidate.  Thus this was a successful \textit{a priori} prediction, and is not a case of \textit{a posteriori\/} reasoning. We follow the evolutionary stages of an SMBHB described by \citet{1980Natur.287..307B}, and assume that the SMBHB has reached the stage where gravitational radiation dominates the energy losses and shrinks the orbit on a timescale $t_\textrm{GR}\propto r^4$, where $r$ is the separation of the SMBHB. It can then be shown that, in observations of duration $\tau_1$, \textit{vs.}~observations of duration $\tau_2$,  the ratio of the numbers of sources in which we might expect to detect strong sinusoidal fluctuations, $N_2/N_1 = [\tau_2/\tau_1]^{(8/3)}$, where $N$ is the number of SMBHBs with separation less than $r$.  We had detected two SMBHB candidates out of 1830 in the OVRO 16-year light curves, and therefore, in the 46-year light curves of the 83 blazars in the combined OVRO+UMRAO, we predicted that we would see strong sinusoidal periodicities in $83\times(2/1830) \times (46/16)^{8/3}  \sim 1.5$ sources. One SMBHB candidate had already been found (PKS 2131$-$021) and therefore there was an expectation of $\sim 0.43$ that a second SMBHB candidate would be found in the combined OVRO+UMRAO sample.  The discovery of the SMBHB candidate PKS J1309+1154 (Papers IV and V) confirmed this prediction.  

While these numbers are small and the uncertainties are therefore large, the numbers of SMBHB candidates found thus far are consistent with the hypothesis that these are SMBHBs. The significance of this result is that it explains why we have not thus far discovered any SMBHB candidates with rest frame periods shorter than a year in the OVRO sample: these shorter period SMBHBs are too rare to have been detected in this sample.  This conclusion will tested as much larger samples of blazars, extending to much weaker flux densities, are observed over the next decade.

\textit{Summary of this section}: The absence of short-period SMBHB candidates \textit{does not\/} constitute evidence for the physical reality of the observed periodicities in the three SMBHB candidates PKS J0805--0111, PKS J1309+1154, and PKS 2131--021, but it \textit{is \/} consistent with the expected relative  distribution of SMBHB orbital periods. This absence of short-period periodicities has to be accounted for in any non-SMBHB model.

\section{Are the periodicities in PKS j0805$-$0111, PKS J1309+1154, and PKS 2131$-$021 real?}\label{sec;real}

In this section we present (i) four arguments \textit{against\/} the hypothesis that some of the periodicities observed in these three SMBHB candidates are real, and (ii) five arguments \textit{for\/} the hypothesis that some, and possibly all, of the periodicities observed in these three SMBHB candidates are real, plus two arguments showing consistency with the SMBHB hypothesis.

\begin{figure*}[!t]
   \centering
   \includegraphics[width=0.9\linewidth]{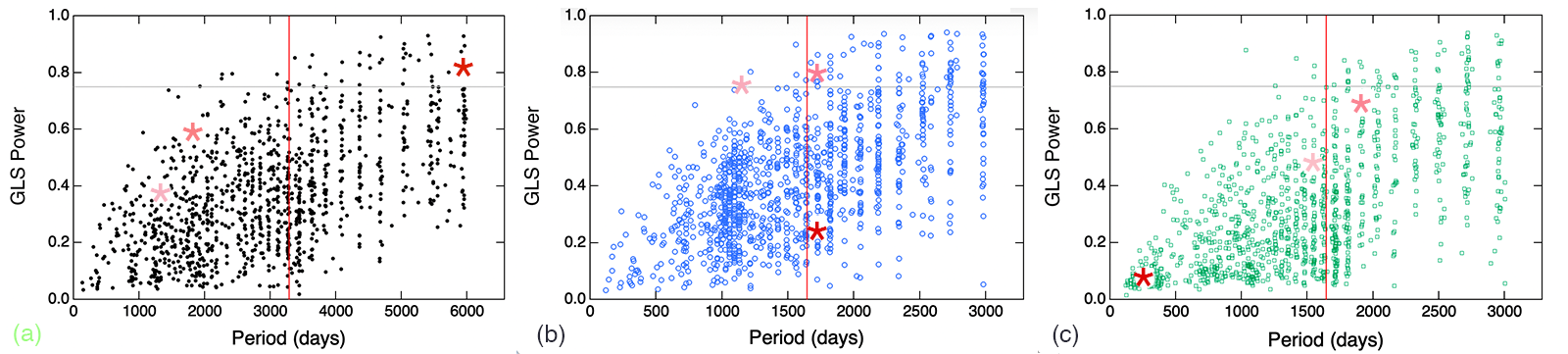}
   \caption{The distributions of the strongest power peaks identified in the GLS periodograms of  1155 sources in the complete sample defined by \citet{2011ApJS..194...29R} (see text). (a) the complete 2008-2025 OVRO 18-year light curves. (b) the 2008-2016 light curves. (c) the 2017-2025 light curves. The light red, medium red, and dark red  asterisks show the results for PKS J0805$-$0111, PKS 2131$-$021, and PKS J1309$+$1154, respectively. The results for PKS J0805$-$0111 and PKS 2131$-$021 show the strongest GLS powers in (b) because of the deviation of these light curves from sinusoidal variations at later times. The varying periods of the strongest power peaks in these three SMBHB candidates over these different observing intervals illustrate the danger of attempting to identify SMBHB candidates based on their GLS power alone.}
   \label{plt:powers}
\end{figure*}

\subsection{The case \textbf{against} some, or possibly all, of the periodicities in PKS J0805$-$0111, PKS J1309+1154, and PKS 2131$-$021 being real}\label{sec:against}

\subsubsection{The Observed Periods are a Significant Fraction of the Duration of the Observations}\label{sec:against1}

It has long been known \citep[see, e.g.,][]{1978ComAp...7..103P} that in astronomical object light curves with a strong red noise component, any apparent periodicities with periods longer than approximately $1/3$  the total duration of the light curve are unlikely to be real, and must therefore be treated with great caution. Furthermore, in Paper IV it was shown that \textit{false\/} periodicities and harmonics, identified by the common variability spectrum  analysis methods (GLS, WWZ, etc.), appear in the vast majority of blazar radio light curves. In the present case, the observed periodicities and the duration of the light curve, during which the periodicities are seen, are given in Table \ref{tab:periodduration}.  The number of periodicity cycles observed in these three SMBHB candidates ranges from 2.5 to 5.5 and therefore are small enough that the possibility that they are random variations caused by the steep slope of the PSD must be taken seriously.

In Fig.~\ref{plt:powers}(a) we show the distribution in power \textit{vs.\/} period of the strongest peak in the GLS periodograms of 1155 of the 1158 sources in the complete sample selected by \citet{2011ApJS..194...29R} that have been observed continuously since 2008.\footnote{Two of the sources in the original sample (J0143+4129 and  J0151-1732) were not observed for the whole period for reasons we are investigating, and are therefore not included in this discussion. A third, J1310+3233, was dropped because of confusion since it is almost exactly 13 arc minutes from a bright blazar, and therefore in the correlated beam of the 40 m Telescope dual-beam system. }  In Figs.~\ref{plt:powers}(b) and (c) we show the same distributions for the first half of the observing period, from 2008-2016, and  for the second half of the observing period, from 2017-2025, respectively. The vertical red lines indicate a period of one half of the total duration --- i.e., 9 years in (a) and 4.5 years in (b) and (c).  The horizontal gray lines indicate GLS power of 0.75. We indicate the GLS power value 0.75 because the powers observed in PKS 2131$-$021, PKS J0805$-$0111, and PKS J1309+1154 were 0.81 (Paper I), 0.77 (Paper III),  and 0.86 (Paper IV), respectively, so they are all above this fiducial value, when the MJD range is restricted to the periods showing sinusoidal variations, as was done in the above papers.

Of interest to this particular discussion is the fact that in (a) six objects have power greater than 0.75 and period less than one half of the total period, meaning that at least two cycles were observed, and in (b) and (c) the number is 5 in both cases. In all three cases the shortest period observed has $3-3.5$ cycles over the duration of the observations.  In all three cases $\sim 0.5\%$ of the objects have powers $> 0.75$ and $2-3.5$ cycles.  Thus, statistically, it could well be the case that the SMBHB candidates that have been identified in the OVRO+UMRAO monitoring programs are simply random periodicities due to the steep slope of the PSD.  The values of the powers and periods of the three SMBHB candidates are indicated by the asterisks in Figs.~\ref{plt:powers}(a), (b), and (c).  Note that these differ from those reported in the above papers because of the cessation of the sinusoid in the case of PKS J0805$-$0111, the fact that the whole 46-yr UMRAO+OVRO light curves was used in the case of PKS J1309+1154, and in the case of PKS 2131$-$021 the light curve has recently departed from the sinusoid.  

These results show that searches for SMBHBs  using the GLS power, while they might help to pick out objects of potential interest, do not provide  a foolproof search method, and any potential SMBHB candidates arising from such searches must be treated with caution.

\subsubsection{Some of the  Periodicities are Sporadic}\label{sec:against2}

As can be seen in both Figs.~\ref{plt:PKS2131harmonic}(a) and  \ref{plt:PKSJ0805harmonic}, the periodicities in PKS J0805$-$0111 and PKS 2131$-$021 disappear abruptly, as they might be expected to do if they are not real periodicities but rather a product of  random variations due to the steep slope of the PSD. Particularly notable in this respect is the dramatic departure from sinusoidal activity seen in PKS 2131-021 in Epoch 4 (Fig. \ref{plt:PKS2131harmonic} (a)).

\subsubsection{Poor fit to the harmonic in PKS J1309+1154 after 2016}\label{sec:against3}

While the evidence for the harmonic in PKS J1309+1154 presented in Paper V is compelling in many respects, it should be noted that the harmonic appears to be much weaker relative to the random correlated variations after 2016.  Thus this putative harmonic is only clearly visible for 4 cycles and is therefore subject to the criticism described above in \S \ref{sec:against1}.

\subsubsection{The Statistical Arguments of Papers I and II}\label{sec:against4}
In Paper I (Table 2) six statistical tests were applied to the data.  The first four of these test are straightforward, and, to the best of our knowledge, correct.  However in the last two tests the p-values are multiplied together, which is not correct, since these correspond to different hypotheses.  Thus these last two tests should be ignored.  Similarly in Paper II, Table 3, the first three tests are straightforward and correct, to the best of our knowledge, but the last test, in which, again, p-values are multiplied, is not correct and should be ignored.  Thus, while the significances of the periodicities reported in Papers I and II are high, in some instances they are not as high as reported in Tests \#3 and \#4 in Table 2 of Paper I,  and in test $1+2+3$ in Table 3 of Paper II. The other p-values in Papers I and II remain unchanged.

\begin{table}
\caption{Observed numbers of cycles}
\label{tab:periodduration}
\centering
\footnotesize
\begin{tabular}{cccc}
\hline \hline
SMBHB&Period &Duration&Cycles\\
Candidate&(yr)&(yr)&\\
\hline
PKS J0805$-$0111       & $3.5$      & 12 &3.4\\
PKS J1309+1154  & 17.9  &46 & 2.5\\
PKS 2131$-$021  & 4.8  & 26& 5.5\\
\hline
\end{tabular}
\tablecomments{The periods are the observed periods and the durations are the duration of the light curve over which sinusoidal variations were seen. Note that in the case of PKS 2131$-$021, as can be seen in Fig.~\ref{plt:PKS2131harmonic}(a), there was a 19-yr gap, from 1984 to 2003, during which the fundamental periodicity was absent. }
\end{table}

\subsection{The case for \textbf{some}, or possibly all,  of the periodicities in PKS J0805$-$0111, PKS J1309+1154, and PKS 2131$-$021 being real}\label{sec:for}

\subsubsection{In-phase reappearance of the fundamental  periodicity in PKS 2131$-$021 after 19 years}\label{sec:for1}

As discussed in Paper I and seen in Fig.~\ref{plt:PKS2131harmonic}(a), a sinusoidal periodicity having period $P=1729.1 \pm 32.4$ days was observed in PKS 2131$-$021 from 1975 to 1983.  This periodicity then disappeared for 19 years, from 1984 to 2003. In 2004 a new periodicity appeared having period $P=1760.4 \pm 5.3$ days. The difference in periods between Epoch 1 and Epoch 3 is thus $\Delta P=30.9 \pm 32.8$ days and therefore not significant.  Furthermore, the phases of these two periodicities differ by only 5\% of a period.  As shown in  Paper I using simulations having the same PSD slope as that observed, the p-value for the Epoch 3 sinusoid being a random red noise fluctuation is $4.0 \times 10^{-4}$, and the p-value for the Epoch 1 period lying in the $\pm 3 \sigma$ window of the Epoch 3 period is $0.020$. Together with the 5\% agreement in  phase, in our view this amounts to a compelling argument that the periodicities observed in Epoch 1 and 3 in PKS 2131$-$021 are real.

\subsubsection{The coherent sinusoidal variations in PKS 2131$-$021 from 2.7 GHz to optical wavelengths }\label{sec:for3}

In Paper II it was shown that PKS 2131$-$021 exhibits strong coherent sinusoidal variations that dominate the light curve, and have the same period, from 2.7 GHz to optical wavelengths, i.e., over a range of $2 \times 10^5$ in wavelength.  Since the emission over this large range originates in different parts of the blazar, it is extremely unlikely that the periodicity is a random result due to the steep slope of the PSD.  The question then arises as to whether the sinusoidal variations might be due to an instability in the relativistic jet.  Here, again, it seems extremely unlikely that that a fluid instability could exhibit such phase coherence over a range of $2 \times 10^5$ in frequency.

\subsubsection{Compelling evidence for harmonic in PKS J1309+1154}\label{sec:for4}

The fundamental and harmonic periodicities in PKS J1309+1154 have been discussed in detail in Paper V, and we summarize them briefly here.

\begin{enumerate}
\item The standard (Paper I) significance test based on simulations shows that the fundamental periodicity is significant at the p-value$\,=\,0.011$ level. 
\item The curvature of the residual light curve after removing the trend and the large amplitude sine wave changes sign, as expected if there is a harmonic term, at each crossing of the putative harmonic and the fundamental --- i.e. 11 times. This has p-value$\,=\,4.9 \times 10^{-4}$.
\item The light curve displays two comparatively flat 8-year-long sections, from 1990 to 1998 and again from 2008 to 2016. The fundamental + harmonic curve likewise displays flat regions over these two 8-year windows, whereas the fundamental curve alone does not.
\end{enumerate}

\subsubsection{Evidence of subharmonics in PKS J0805$-$1154 and PKS 2131$-$021}\label{sec:for5}

The evidence for the 2:1 subharmonics in PKS J0805$-$0111 and PKS 2131$-$021, which contribute a significant fraction of the power in the variability spectrum, as revealed through the ratio of the rms deviations, while not yet compelling, is strong enough that these should be considered as likely to be real, and in our view it would be a mistake  to ignore them.

\subsubsection{Evidence for both periodicities in PKS J0805$-$1154 and PKS 2131$-$021 being real}\label{sec:for6}

A strong argument for the fundamental and subharmonic periodicities in PKS J0805-0111 and PKS 2131-021 being real is the fact that the fundamental and the subharmonic in PKS J0805-0111 disappear simultaneously, and in PKS 2131-021 the fundamental disappears abruptly at the end of Epoch 1 and the subharmonic appears abruptly at the beginning of Epoch 2, i.e., immediately following the disappearance of the fundamental.

If these harmonics and subharmonics were random sinusoidal variations due to the steep slope of the PSD then  they would not be expected to appear/disappear simultaneously.

\subsubsection{Consistency Argument: The Absence of short period SMBHB candidates}\label{sec:for7}

While consistency with the expected distribution of periods is not an argument \textit{for\/} the observed periodicities being real, it is clearly an important test of any model that it should be consistent with the statistics, as far as they are known.  The statistics are discussed in \S \ref{sec:incidence}. On the SMBHB hypothesis one expects far more long-period binaries than short-period binaries due to the rapid decay of the orbit once gravitational radiation becomes significant --- so the current statistics, though small, do not refute the hypothesis that these periodicities are real. 

While the above gravitational radiation argument applies to SMBHBs, it does not apply to binaries in which the secondary has mass much less than the binary.  But we note that it is probably generally true that provided the evolution times are short compared with the age of the source, the lifetimes will be shorter at shorter periods.

\subsubsection{Consistency Argument: The large flare in PKS 2131$-$021 in Epoch 4}\label{sec:for8}
While the argument against these periodicities being real presented  in \S \ref{sec:against2} should be taken seriously, the flare in Epoch 4 shows a completely different dependence on frequency than is seen in the sinusoidal variations (Paper II). An analysis of this flare by Diamantopoulos et al. (in preparation) shows that the lag between the flares at 15 GHz and at 91.5 GHz is $\ll 127$ days, the lag between these two frequencies  that applies to the sinusoidal variations (Paper II).  It is therefore clear that the flare emission comes from a different region within the core than the region producing the sinusoidal variation.  The evidence suggests that the sinusoidal variations in PKS 2131--021 are  present during the flare, but heavily masked by it.

\section{Requirements of The Models}\label{sec:requirements}

The phenomenology that we wish to explain in our models is the following:
\vskip 6pt
\noindent
I. The three blazars discussed in this paper all show strong evidence of real periodicities, making them strong SMBHB candidates.
\vskip 6pt
\noindent
II. In all three there are two clear periodicities that have a 2:1 periodicity relationship, either a fundamental and a first harmonic (PKS J1309+1154) or a subharmonic and a fundamental (PKS J0805-0111, PKS 2131-021).
\vskip 6pt
\noindent
III. In two of the sources the fundamental and harmonic/subharmonic are visible at the same time (PKS J0805-0111, PKS J1309+1154).
\vskip 6pt
\noindent
IV. In one of the sources the fundamental and subharmonic are not visible at the same time (PKS 2131-021).
\vskip 6pt
\noindent
V. In one of the sources the fundamental and subharmonic disappear at the same time (PKS J0805-0111).
\vskip 6pt
\noindent
VI. In one of the sources the fundamental and harmonic are in phase to high precision (PKS J1309+1154).
\vskip 6pt
\noindent
VII. In two of the three sources the fundamental and subharmonic is not (PKS J0805-0111), or may not be (PKS 2131-021), in phase.
\vskip 6pt
\noindent
VIII. The absence of candidates having periods shorter than a few years  has to be accounted for in any successful model.

\section{Models}\label{sec:modeling}

\subsection{Modified Kinematic Orbital (MKO) Model}\label{sec:MKOmodel} 
One of the simplest interpretations for the observed fundamental periodicities in these three blazar light curves is modulation of the relativistic jet by the orbital motion of the SMBH emitting the jet \citep[][Papers I-VI]{2017MNRAS.465..161S}. The model presented in Paper I and \cite{2017MNRAS.465..161S}, which we call the ``kinetic orbital'' (KO) model, is a multi-zone model in which the jet is carried around in parallel transport by one of the black holes and propagates ballistically. In the modified kinematic orbital (MKO) model, detailed in Paper VI, the jet is guided by a wind: this is a multi-zone model in which one black hole contains an accretion disk and launches both a jet and a wind. The wind, dragged around by the orbital motion, creates a channel through which the relativistic flow travels. A key motivation for invoking the wind is to get the propagation of the jet pattern to ``slow down'' so that one can explain the evolution of the phase with observing frequency at reasonable distances from the jet base, i.e.~consistent with the Blandford-K\"onigl model \citep{1979ApJ...232...34B}.  

The fundamental periodicities in PKS 2131, J0805, and J1309 have all been interpreted in the context of the MKO model (Papers IV, VI). 

We now turn to the harmonic periodicities in these three objects. In the case of PKS J1309+1154, Paper IV invoked a two-jet model to explain the harmonic, but as we discuss below, the harmonic in PKS J1309+11154 can be explained on the one-jet MKO model.  We first consider the possible harmonics in PKS J0805--0111 and PKS 2131--021 found in this paper. The harmonics in these two objects challenge the most basic MKO model as applied to SMBHBs. In particular, in the case of PKS 2131-021, the appearance and disappearance of the sub-harmonic in Epoch 2, if real, would be rather challenging to explain. This led us to consider models in which the secondary is not an SMBH, but much less massive, and possibly not a black hole. We discuss this in \S\ref{sec:lessmassive}.

\subsubsection{Single Jet Models}
The model put forward in Paper VI is that one of the two black holes launches a wind-collimated jet, while the other is dormant. The so-called mini-disks around component black holes in the binary are thought to be fed by a circumbinary disk \citep{2017ApJ...838...42B, 2018ApJ...853L..17B, 2024ApJ...973L..19M, 2025ApJ...979L..24R}. For only one strong jet to appear, the feeding from the circumbinary disk would need to concentrate on one of the black holes (most likely the primary), giving it a stronger mini disk to launch and collimate the jet. This was the model applied to PKS 2131--021 and PKS J0805--0111 in Paper VI. 

The single jet model naturally creates harmonics (although this was neglected in Paper VI), since the Doppler factor which beams the emission is a non-linear function of the jet velocity. The jet opening angle and viewing inclination angle both affect the relative strength of higher harmonics. Jets with large opening angles show less sharp peaks in their light curves, and hence weaker harmonics,  because more of the jet is visible to an observer at a given time. Smaller opening angles require smaller inclination angles between the jet axis and viewer line of sight, leading to sharper peaks, and hence stronger harmonics. Both the number of visible emitting regions associated with the opening angle of the jet and the viewing inclination have an effect on harmonics which naturally result from integrating the Doppler boosted emissivity over the whole jet (Paper IV). The sharper peaks in PKS J0805--0111 as well as the harmonic in PKS J1309+1154 could be due to these higher harmonic contributions. A quantitative study of this model is left to future work.

If real, the subharmonic in Epoch 2 of PKS 2131 complicates the single jet story since this is definitely not seen in Epochs 1 and 3. Under a single jet model, the Epoch 2 light curve could reflect a change in the fueling of the jet or, alternatively, the integrity of the collimating wind. Both of these cases would result from changes to the conditions of the jet-launching black hole's mini-disk, and by extension the feeding from the circumbinary disk, which can be quasi-periodic, with periods related to the orbital period of the circumbinary disk inner edge \citep[e.g.][]{2018ApJ...853L..17B}. Thus, the subharmonic in Epoch 2 could be a quasi-periodic oscillation related to the longer orbital period of the inner edge of the circumbinary disk. 
A resonance between the binary orbit and the circumbinary disk might cause the circumbinary disk inner edge to have a period which is a multiple of the binary period $P_b$, particularly if the binary is somewhat eccentric \citep{2008ApJ...672...83M, 2012ApJ...749..118S, 2012ApJ...755...51N, 2026arXiv260620787G}. Recent simulations have shown that this inner feature can have a period of of $4P_b$, which is imprinted on the rate of accretion onto the binary \citep{2026arXiv260620787G}. As a consequence, a strong harmonic also appears at $2P_b$, as we observe in Epoch 2 \citep{2026arXiv260620787G}. The $4P_b$ periodicity is just barely too long to be observed in our Epoch 2 baseline, but would account nicely for the long-term trend seen in Epoch 2 mentioned above in \S \ref{sec:2131} and in the Appendix. These periods could then be imprinted on the observed jet emission by causing the jet power to fluctuate with the periodic mini-disk feeding.

To summarize, in the single jet interpretation of PKS 2131, the period in Epochs 1 and 3 is the binary orbital period. This would explain why the sinusoid appears so stable in these epochs. The Epoch 2 period is associated with the inner circumbinary disk and, while less stable, may reflect a resonance with the binary orbit. We predict that if the source undergoes a ``future Epoch 2" and the subharmonic returns, it will not necessarily be phase coherent with the observed Epoch 2, while a ``future Epoch 1/Epoch 3" should be phase coherent with the previously observed Epochs 1 and 3.

\subsubsection{Dual Jet Models}
In Paper IV, we interpreted the harmonic in J1309+1154 as possibly coming from two jets --- that is, both black holes in the SMBHB each launch their own jet. If the jets are unequal in power, the periodic light curve will contain two peaks out of phase by half a period, with one peak brighter than the other. While a viable model for J1309, there remain physical problems with this model. For example, for a double jetted source to be a blazar, both jets must be pointed in the same direction. For this to happen the black hole spins would need to be aligned, which could happen due to angular momentum exchange with the circumbinary disk. Nonetheless, even two parallel jets can tangle each other up as they snake around one other, disrupting their collimation \citep{2025ApJ...979L..24R}. If the twisting jets collide upstream of the emission region, we would not expect clean sinusoidal variations.

If applied to PKS 2131, the dual jet model could have implications for the Epoch 2 subharmonic. Under this interpretation, both black holes in the SMBHB would launch powerful co-equal jets during Epochs 1 and 3. The 2.1 yr source-frame period of Epochs 1 and 3 would then be half the period of the SMBHB. The change from Epoch 1 to Epoch 2 would then be due to one of the two jets turning off. The period seen in Epoch 2 would then be the actual SMBHB period as only one jet produces sinusoidally varying emission. In Epoch 3, the second jet then returns. Why the jets would have equal powers during Epochs 1 and 3, but not during Epoch 2 is not clear. Moreover, what causes the average flux and sinusoidal variation amplitude to differ between Epochs 1 and 3 is not obvious. 

Alternatively, if neither jet switches off during Epoch 2, then under the dual jet model, the transient sub-harmonic in Epoch 2 could be caused by the precession of one of the jets.  During Epoch 2, one jet may precess away from the viewer, leaving only the other jet in view. A prediction of this model is that the height of one peak should vary over time as the jet comes in and out of view on the timescale of the precession. Indeed, we may be seeing this in the peak at MJD $\sim 51800$. The precession timescale would have to be a few orbital periods to explain the observations.

\subsection{Less Massive Secondary}\label{sec:lessmassive}
In this section we discuss magnetospheric modulation by a less massive orbiting companion.  The jet modulation is associated with a low mass secondary, $M_S$, that does not appreciably move the primary, with mass $M_P \gg M_S$, but does modulate the magnetohydrodynamic (MHD) wind.

We suppose the orbit of the secondary has been circularized through drag imposed by the surrounding gas so that its radius is $r/10^{16}\,{\rm cm}\equiv r_{16}\sim1.5P_y^{2/3}M_{P9}^{1/3}$ and its velocity (from Kepler's Laws) is $v/10^9\,{\rm cm\,s}^{-1}\equiv v_9\sim3P_y^{-1/3}M_{P9}^{1/3}$, where $M_{P9}$ is the mass of the primary in units of $10^9\,{\rm M}_\odot$. If we suppose that the blazar jet is powered by black hole spin and confined by an MHD wind launched by an orbiting disk, a simple (and quite uncertain) theoretical model leads to an estimate of the magnetic field strength in the wind  at a radial distance $r=10^{16}r_{16}\,{\rm cm}$ from the primary of $B_W\sim L_{j\,44}^{1/2}r_{16}^{-1}\,{\rm G}$, where $L_j=10^{44}L_{j\,44}\,{\rm erg\,s}^{-1}$, is the jet power. This estimate is based upon simple models of magnetocentrifugal winds from accretion disks where the strength of the magnetic field scales very roughly inversely with radius \cite[e.g.][]{1997ApJ...486..291O}. 

The secondary might be a massive, Wolf--Rayet-like star with mass $M_S\sim100\,{\rm M}_\odot$ accompanied by a wind with mass loss $\sim10^{-5}\,{\rm M}_\odot\,{\rm y}^{-1}$ and speed $\sim3000\,{\rm km\,s}^{-1}$ in an inclined orbit around the primary. Such stars are very rare in our Galaxy, with roughly a thousand observed,  but might be much more common in the central SMBH-dominated regions of galaxies \citep{2015MNRAS.447.2322R}. The wind speed is much less than the orbital speed. The gas that is lost by the star is likely to orbit the primary under a combination of gravitational and magnetic force. We can estimate the radius $r_\ast$ from the star at which the stellar wind momentum flux balances the magnetic pressure in the primary disk wind and obtain: $r_\ast/r\sim0.2L_{j44}^{1/2}M_{P9}^{1/3}P_y^{-1/3}$. 

As mentioned in the Introduction, any massive star on a trajectory that takes it close enough to the SMBH to create a TDE might possibly evolve into a CSO, as described in \citet{1994cers.conf...17R}, \citet{2024ApJ...961..242R}, and \citet{2024MNRAS.528.6302S}, which might explain the absence of periodicities shorter than $\sim 3$ years amongst the OVRO sample of $\sim 1830$ blazars.

Alternatively, the secondary might be an intermediate mass black hole (IMBH) with a small accretion disk and wind outflow. The region around this structure is plausibly of similar size. In both examples the secondary is responsible for a significant perturbation to the wind and, consequently, to the primary, blazar jets.

A third possibility is a young pulsar or a magnetar, like SGR J1745$-$2900, \citep{mori13}, which orbits Sgr A$^\ast$ with an orbital period of a few thousand years. Approximately $10^{47-48}\,{\rm erg}$ of essentially electromagnetic energy could plausibly be released into a disk magnetosphere and, as with a massive star wind, could modulate a blazar jet periodically. A young pulsar/magnetar would likely be created in an inclined orbit through a natal kick.

Next, we consider how the secondary might influence the motion of and emission from the single pair of oppositely-directed jets from the primary massive black hole. The MHD wind outflow from the disk orbiting the primary black hole can be divided into a low latitude, sub-Alfv\'enic region that is in causal contact through the disk, and a high-latitude, super-Alfv\' enic region. 

Schematically, when the secondary passes through the disk, it can strongly perturb the magnetosphere on both sides of the disk. This will happen twice per orbital period\footnote{What is being proposed has some similarity to a likely explanation \citep{dai13} of Quasi-Periodic Eruptions \citep{kara25} where a lower mass star induces a burst of X-ray emission whenever it crosses the disk. The periods are smaller --- days as opposed to years --- than those discussed here and are associated with lower mass main sequence stars or even white dwarfs.}. It is not clear how the wind from the orbiting star modulates the jet as it orbits its source, presumably a spinning black hole. If it changes the jet direction, then we might expect the response in the observed flux to have opposite signs on opposite sides of the orbit. In this case, the observed fundamental would, instead, equal the orbital period. Alternatively, the response may be to change the jet speed, giving an observed fundamental equal to half the orbital period. The phase difference of either modulation expected from the disk crossing is very hard to predict.

If the inclination of the orbit is large, so that the star spends some of each orbit in the super-Alfv\'enic part of the wind, its influence will be confined to one jet and the associated period in the approaching jet flux will exhibit the orbital period. Again, the nature of the modulation is model-dependent and can either change the angle or the speed. If the orbital inclination is low, there will be no such modulation. It should be possible to perform MHD simulations to check on some of these speculative assertions. 

Now, turn to the behaviors discussed in \S\ref{sec:requirements}. The fundamental and the subharmonic may be visible at the same time. This is behavior III, reported for PKS J0805-0111 and PKS J1309+1154, either as a result of a complex response to disk crossing that includes variation of the jet direction and speed, or as a combination of a jet speed modulation with a high latitude interaction. If the fundamental and the  subharmonic are not exhibited simultaneously --- behavior IV, as exhibited by PKS 2131-021 --- then this might indicate switching from speed modulation to direction modulation reflecting changes in the nature of the disk flow. A dramatic reduction in the disk flow could lead to a loss of all modulation (behavior V). If the fundamental and subharmonic are in phase, this could indicate that the observer direction is very close to the jet axis, with the true fundamental being the identified subharmonic and the observed variation being time-symmetric. 

The model just sketched is quite fragile and there are several ways that it could be effectively ruled out. The massive stars and/or IMBHs should be very good clocks over the timescales of observation. There could be considerable variation in the phase of the modulation but not in the underlying periods. Unless IMBHs and massive stars are more common than expected, it would be surprising to see several incommensurate large periods in a single source. It seems more reasonable to suppose that orbiting low mass stars, including red giants, are much  more common and that shorter periods will be detected in the future.

Perhaps the most convincing way to rule out models in which the modulation of a blazar jet is mediated indirectly through an enveloping MHD wind rather than directly through gravitational force acting on its black hole source would be to find evidence of additional disks or jets associated with one or more secondary black holes. Careful spectroscopic or polarimetric investigations of the three sources under consideration here are well-motivated.


\subsection{SMBH Satellite Systems}\label{sec:satellite}
A simple model arises from analogy to the Galilean moons of Jupiter or to young extrasolar planet systems. The three innermost Galilean moons, Io, Europa, and Ganymede have orbital periods in the ratio of 1:2:4, the result of mean motion resonances; mutual perturbations between the satellites maintain the period ratios in the face of external non-Keplerian forces, in this case tidal interactions with Jupiter. Similarly, many extrasolar planet systems with two or more planets are seen to be in or, more commonly, near mean motion resonances, particularly in young systems \citep{2024AJ....168..239D}. In the case of exoplanets, the putative resonant configurations are attributed to inward migration of planets in the protostellar disk out of which they formed.

The situation in active galactic nuclei (AGN) resembles that in protostellar disks; a central massive object (a supermassive black hole in an AGN) is surrounded by a gaseous disk, which may contain stars or black holes. By analogy with protostellar disks, if there are two black holes in orbit around the more massive central black hole, interactions with the accretion disk may push the stars or black holes toward each other, into a mean resonance\citep[e.g.,][]{2016ApJ...819L..17B}. The first strong resonance encountered by converging objects, assuming they start fairly far apart, is the 2:1 mean motion resonance. Depending on the mass of the two converging bodies, and on the properties of the disk, they may be trapped into the 2:1 resonance; for strong disk induced torques, the migration may continue, leading the migrators to either be trapped into or pass through the 3:2 resonance, the  4:3 resonance, and so forth.

Once trapped into a resonance, the two bodies provide a clock, with two strong harmonics (as well as, potentially, higher harmonics). Mechanisms similar to those described in \S \ref{sec:lessmassive} could then affect the jet emerging from the central supermassive black hole.

\section{Discussion}\label{sec:discussion}

It is well known that false periodicities lasting for just a few cycles are common in blazars \citep{1978ComAp...7..103P,2016MNRAS.461.3145V}.  Indeed \citet{1978ComAp...7..103P} cautions specifically against periodicities that are not statistically real that have periods of about one third of the duration of the observations. It should be noted that the three strong SMBHB candidates identified thus far in the combined OVRO+UMRAO monitoring samples are in the danger zone, with 3.4 cycles for PKS J0805$-$0111, 2.5 cycles for PKS J1309+1154, and 5.5 cycles for PKS 2131$-$021.  

In addition, while the GLS spectra presented in Papers I, II, and III are encouraging, these are not sustained in  the full light curves of PKS J0805$-$0111 and PKS 2131$-$021  up to the end of 2025. In the case of PKS J1309+1154 the light curve is predicted not to show sinusoidal variations from 2026-2034, after which the 17.9 year periodicity is predicted to resume. 

The statistical reality of these three SMBHB candidates therefore has to rest on more than the apparent periodicities alone. This led to the extensive statistical tests using simulations with matching PSD slopes and PDF distributions in Papers I--IV and the search for harmonics in Paper V and in this paper.  All of these three SMBHB candidates pass stringent simulation and harmonic tests.  In addition, as described in Paper II, the sinusoidal variations seen in PKS 2131$-$021 extend coherently to optical wavelengths, in spite of the fact that the emission from these regions cannot be closely connected and must be due to an underlying physical periodicity that is responsible for both.

Our conclusion is that there is strong evidence for statistically real periodicities in all three of these SMBHB candidates, so they merit exhaustive study. Independent evidence, e.g.~optical observations of periodic variations in spectral lines that enable the determination of orbital velocities, and/or observations of gravitational waves with  millisecond pulsar arrays, would place the statistical reality of these periodicities beyond doubt.

\section*{ACKNOWLEDGEMENTS}
This work is supported by NSF grants AST2407603 and AST2407604. We thank the California Institute of Technology and the Max Planck Institute for Radio Astronomy for supporting the  OVRO 40\,m program under extremely difficult circumstances over the last 8~years in the absence of agency funding. Without this private support these observations could not have been made.  We thank all the volunteers who have enabled this work to be carried out. 
Prior to~2016, the OVRO program was supported by NASA grants \hbox{NNG06GG1G}, \hbox{NNX08AW31G}, \hbox{NNX11A043G}, and \hbox{NNX13AQ89G} from~2006 to~2016 and NSF grants AST-0808050 and AST-1109911 from~2008 to~2014.
I.L  was funded by the European Union ERC-2022-STG - BOOTES - 101076343. Views and opinions expressed are however those of the author(s) only and do not necessarily reflect those of the European Union or the European Research Council Executive Agency. Neither the European Union nor the granting authority can be held responsible for them.
W.M. acknowledges support from ANID project Basal FB210003. A.S. and R.B. acknowledge support by a grant from the Simons Foundation (00001470,RB,AS). A.S. thanks K. Chatziioannou, S. Phinney, J. Fuller, M. Garg, and A. Galishnikova for useful discussions. R.R. and B.M. and P.V.d.l.P. acknowledge support from ANID Basal AFB-170002, from Núcleo Milenio TITANs (NCN2023\_002), and CATA BASAL FB210003. P.V.d.l.P. also acknowledges support by the National Agency for Research and Development
(ANID) / Scholarship Program / Doctorado Nacional/2023--21232103.
T.H. acknowledges support from the Academy of Finland projects 317383, 320085, 322535, and 345899.
This paper makes use of the following ALMA data: ADS/JAO.ALMA\#2011.0.00001.CAL. ALMA is a partnership of ESO (representing its member states), NSF (USA) and NINS (Japan), together with NRC (Canada), NSTC and ASIAA (Taiwan), and KASI (Republic of Korea), in cooperation with the Republic of Chile. The Joint ALMA Observatory is operated by ESO, AUI/NRAO and NAOJ. The National Radio Astronomy Observatory is a facility of the National Science Foundation operated under cooperative agreement by Associated Universities, Inc. This research has made use of the NASA/IPAC Extragalactic Database (NED), which is funded by the National Aeronautics and Space Administration and operated by the California Institute of Technology.

--------------------------------------------------------------
\bibliography{references}{}
\bibliographystyle{aasjournalv7}


\appendix

\section{Epoch 2 Phase Uncertainty Including Period Uncertainty and Correlated Noise}\label{sec:corrnoise}

We follow the procedure of Appendix B of Paper II to fit the Epoch 2 light curve. We model the light curve as a standard sinusoid with a linear term to account for the increasing trend observed in the light curve:
\begin{equation}
S(t)=A\sin\left(\frac{2\pi(t-t_0)}{P_1} + \phi_0\right) + S_1(t-t_0) + S_0,
\end{equation}
where $S_1$ and $S_0$ are the coefficients of the linear trend and $t_0=57000$. We perform a preliminary linear trend--only fit to the Epoch 2 light curve using \texttt{numpy} \texttt{polyfit} of degree one and use the resulting parameters and uncertainties as priors for the MCMC fitting with $\log(S_1){\sim}N(-4.10, 0.05)$ and $S_0{\sim}N(-2.22, 0.08).$ We adopt uniform priors on the amplitude and phase $A{\sim}U(0.1, 1)$,  $\phi{\sim}U(0, 2\pi)$. We keep the lower bound of 0.1 on the amplitude to prevent the linear component dominating the fit, and subtraction of the initial linear component shows an amplitude $>0.1\,\mathrm{Jy}$. We keep a normal prior on the period $P{\sim}N(3479.6, 34.8)$ days to test the phase alignment.

\begin{figure}[htp!]
    \centering
    \includegraphics[width=0.7\linewidth]{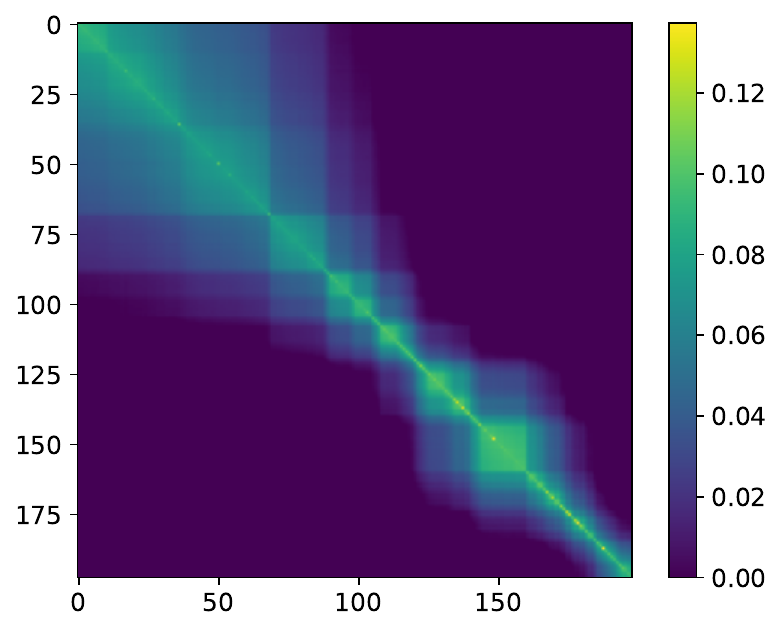}
    \caption{Sample-sample covariance matrix between timesteps across the 10,000 simulations.}
    \label{fig:figA1}
\end{figure}

\begin{figure}
    \centering
    \includegraphics[width=\linewidth]{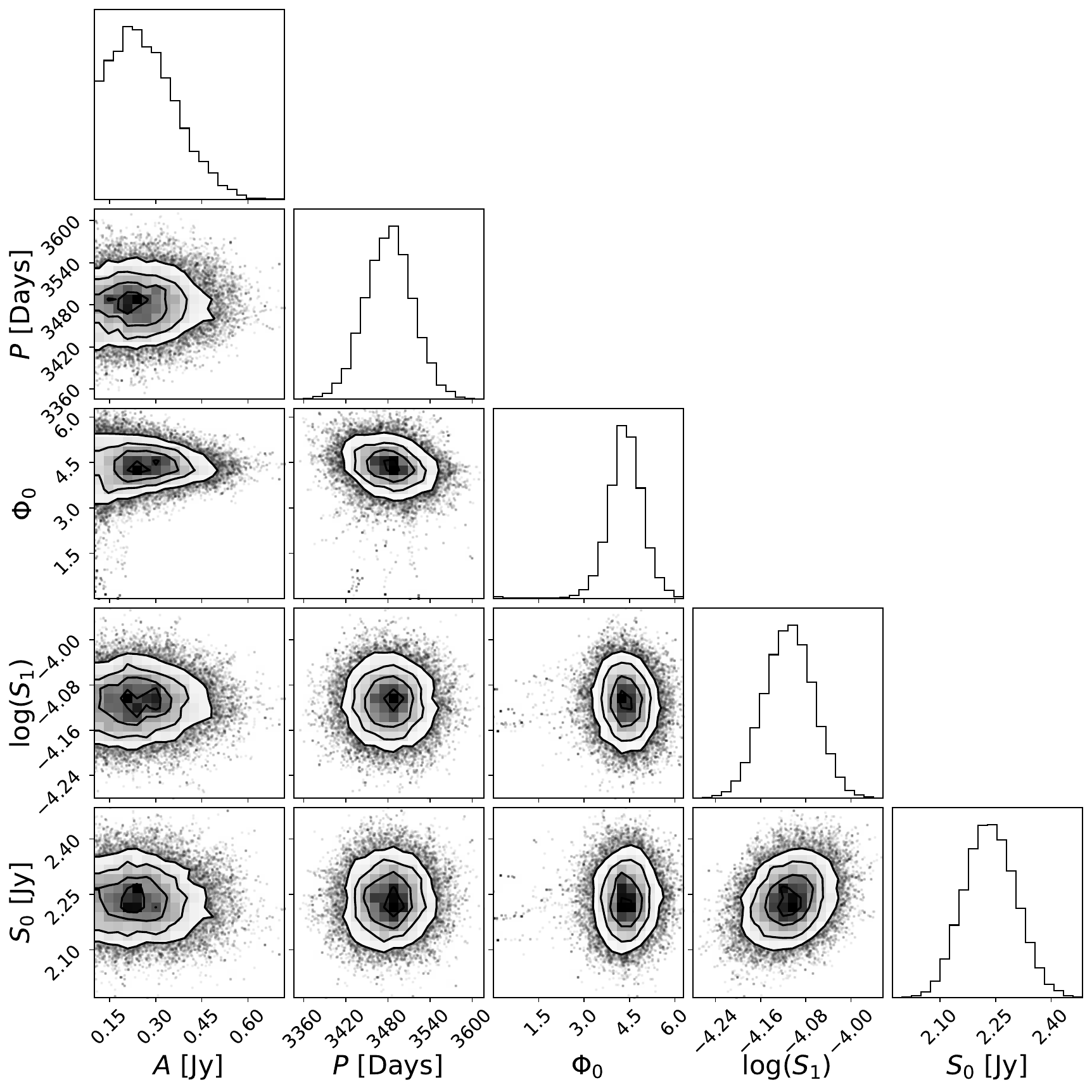}
    \caption{Corner plot showing the posterior of Epoch 2 sinusoid plus linear trend fit.}
    \label{fig:figA2}
\end{figure}

\begin{figure}
    \centering
    \includegraphics[width=\linewidth]{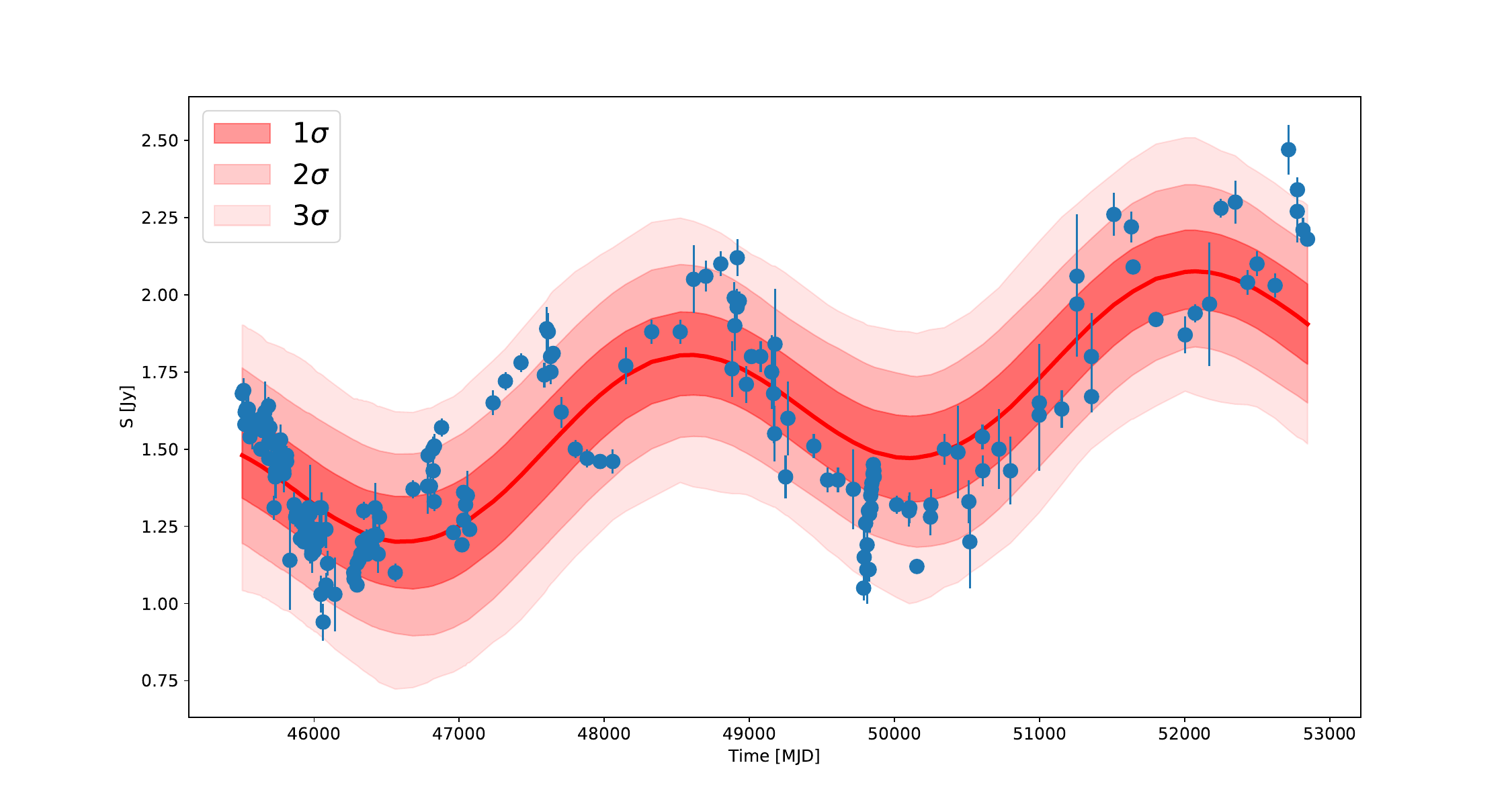}
    \caption{Epoch 2 light curve overlaid with the 1, 2, and 3${\sigma}$ intervals and the median (solid red).}
    \label{fig:figA3}
\end{figure}

We use a multivariate Gaussian log--likelihood as described by Equation B4 in Paper II, where we use a sample--sample covariance matrix estimated from 10,000 simulations sampled over the Epoch 2 time grid using the \cite{2013MNRAS.433..907E} method. A manual 30\% shift is implemented to minimize correlations between measurements at the beginning and end of Epoch 2. However, at sufficiently large timestep separations, a negative covariance develops and starts to rise again, which we set to zero as done by Paper II. The resulting sample-sample covariance matrix is shown in Figure \ref{fig:figA1}. The corner plot of the resulting fit is shown in Figure \ref{fig:figA2}, and 1, 2, and 3$\sigma$ intervals generated from 10000 random posterior samples overlaid with the median of the samples are shown in Figure \ref{fig:figA3}.

\section{Sine-wave fits}\label{sec:sinefits}

 We fitted the following model to the data collected in the time range $54474 \leq \mathrm{MJD} \leq 59041$:
\begin{equation}
S(t) = A_1 \sin\left(\frac{2\pi(t-t_0)}{P_1} - \phi_1\right) + A_2 \sin\left(\frac{2\pi(t-t_0)}{P_2}-\phi_2\right) + S_0,
\end{equation}
where ($P_1$, $A_1$, $\phi_1$) and ($P_2$, $A_2$, $\phi_2$) represent the period, phase, and amplitude of the subharmonic and the fundamental, respectively, $S_0$ is the mean flux density, and $t_0 = 57000$. The best-fit parameters were found by maximizing the following log-likelihood function:
\begin{equation}
\ln \mathcal{L} = -\frac{1}{2} \sum_{i=1}^N \left(\frac{(S_i - S(t_i))^2}{\sigma_i^2+\sigma_0^2} + \ln\left(\sigma_i^2 + \sigma_0^2\right)\right),
\end{equation}
where $S_i$ and $\sigma_i$ refer to the flux density and its uncertainty measured at time $t_i$, while $\sigma_0$ is a parameter that quantifies the intrinsic noise in the light curve of the blazar. We adopted uniform priors on all parameters, and used the \textsc{emcee} sampler \citep{2013PASP..125..306F} to derive their posterior distributions.

The results of the fits are presented in Table~\ref{tab:J0805}. For each model, we also provide the Bayesian information criterion (BIC) score defined as:
\begin{equation}
\mathrm{BIC} = k \ln(n) - 2\ln \mathcal{L}_{\rm max},
\end{equation}
where $n=550$ is the number of data points, $k$ is the number of the parameters of the model, and $\ln \mathcal{L}_{\rm max}$ is the maximum value of the log-likelihood function.

In model~1, we fit only the fundamental component, neglecting the subharmonic by setting $A_2 =0$. The best-fit parameters are essentially identical to those found by \citet{2025ApJ...987..191D}. In models 2 and 3, we fixed $P_2 = 2P_1$. In model~2, the subharmonic has the same phase as the fundamental ($\phi_1 = \phi_2$), and we could provide only a 95\% upper limit on $A_2$. In model~3, we allowed both $A_2$ and $\phi_2$ to vary. Finally, in model~4, we permitted $P_2$ to vary as well.

\begin{table*}[!b]
\caption{Best-fit sine-wave model parameters}
\label{tab:J0805b}
\centering
\footnotesize
\begin{tabular}{lcccc}
\hline \hline
& Model 1 & Model 2 & Model 3 & Model 4\\
\hline
$P_1$ (d)       & $1240.7 \pm 4.5$      & $1240.8 \pm 4.5$      & $1222.1 \pm 3.1$      & $1220.2 \pm 3.1$\\
$A_1$ (Jy)      & $0.1709 \pm 0.0043$   & $0.1709 \pm 0.0043$   & $0.1724 \pm 0.0031$   & $0.1736 \pm 0.0030$\\
$\phi_1$ (rad)  & $-0.142 \pm 0.024$    & $-0.141 \pm 0.024$    & $-0.154 \pm 0.017$    & $-0.166 \pm 0.017$\\
$P_2$ (d)       & \dots                 & $2P_1$                & $2P_1$                & $2541.6 \pm 36.1$ \\
$A_2$ (Jy)      & \dots                 & $<0.0008$             & $0.0679 \pm 0.0030$   & $0.0677 \pm 0.0028$\\
$\phi_2$ (rad)  & \dots                 & \dots                 & $2.974 \pm 0.046$     & $3.017 \pm 0.048$\\
$S_0$ (Jy)      & $0.4294 \pm 0.0030$   & $0.4295 \pm 0.0030$   & $0.4305 \pm 0.0021$   & $0.4320 \pm 0.0021$\\
$\sigma_0$ (Jy) & $0.0662 \pm 0.0022$   & $0.0664 \pm 0.0022$   & $0.0450 \pm 0.0016$   & $0.0446 \pm 0.0016$\\
\hline
BIC             & $-2354.2$             & $-2347.8$             & $-2731.4$             & $-2733.9$ \\
\hline
\end{tabular}
\end{table*}




\end{document}